\documentclass[twocolumn,twoside,slac_two]{revtex4}
\usepackage{graphicx}
\usepackage{fancyhdr}
\def\EPJ                     {\mbox{Eur. Phys. J.}}

\def\NuclPhys                {\mbox{Nucl. Phys.}}

\def\PhysLett                {\mbox{Phys. Lett.}}

\def\PhysRev                 {\mbox{Phys. Rev.}}
\def\PRL                     {\mbox{Phys. Rev. Lett.}}

\newcommand{\Journal}[4]     {{#1} {\bf {#2}}, {#3} ({#4})}

\newcommand{\arXiv}[1]       {{\tt arXiv:{#1}}}

\newcommand{\WWWAddr}[1]     {{\tt {#1}}}

\newcommand{\TwoColFig}[1]{%
  \includegraphics[width=\linewidth,clip=false]{#1}\vspace{0pt}
}

\newcommand {\unitexp}[2] {\mbox{{#1}$^{\mathrm{#2}}$}}
\newcommand {\scinot}[2]  {\mbox{{#1}$\times$10$^{\mathrm{#2}}$}}
\newcommand{\etal}      {\mbox{\it et al.}}
\newcommand {\ra}       {\mbox{$\rightarrow$}}
\newcommand {\Dzero}    {\mbox{D0}}
\newcommand {\Order}[1] {\mbox{$\mathcal{O}({#1})$}}
\newcommand {\CP}       {\mbox{$CP$}}
\newcommand {\Vckm}[1]  {\mbox{$V_{#1}$}}
\newcommand {\Vckmst}[1] {\mbox{$V^*_{#1}$}}
\newcommand {\dderiv}   {\mbox{$\mathrm{d}$}}

\newcommand {\Deltamq}[1] {\mbox{$\Delta m_{#1}$}}
\newcommand {\dmd}        {\Deltamq{d}}
\newcommand {\dms}        {\Deltamq{s}}
\newcommand {\eDsq}       {\mbox{$\varepsilon D^2$}}

\newcommand {\qrk}[1]   {\mbox{${#1}$}}

\newcommand {\ppb}      {\mbox{$p\bar{p}$}}

\newcommand {\Jpsi}     {\mbox{$J/\psi$}}

\newcommand {\Ds}       {\mbox{$D_s$}}

\newcommand {\Bhad}     {\mbox{$B$}}
\newcommand {\Bz}       {\mbox{$B^0$}}

\newcommand {\Bzq}[1]   {\mbox{$B^0_{#1}$}}
\newcommand {\Bzqbar}[1]{\mbox{$\bar{B}^0_{#1}$}}

\newcommand {\Bd}       {\mbox{$B_d$}}
\newcommand {\Bs}       {\mbox{$B_s$}}
\newcommand {\Bc}       {\mbox{$B_c$}}

\newcommand {\Bp}       {\mbox{$B^+$}}

\newcommand {\Lb}       {\mbox{$\Lambda_b$}}

\begin{document}

\title{B Lifetimes and Mixing}

\author{H. Evans}
\affiliation{Indiana University, Bloomington, IN, USA --
  for the CDF and \Dzero\ Collaborations}

\begin{abstract}
The Tevatron experiments, CDF and \Dzero , have produced a wealth of
new \Bhad -physics results since the start of Run II in 2001. 
We've observed new \Bhad -hadrons, seen new effects, and increased
many-fold the
precision with which we know the properties of \qrk{b}-quark systems.
In these proceedings, we will discuss two of the most
fruitful areas in the Tevatron \Bhad -physics program: lifetimes and
mixing. We'll examine the experimental issues driving these analyses,
present a summary of the latest results, and discuss prospects for the
future. 
\end{abstract}

\maketitle

\thispagestyle{fancy}

\section{\label{sect:intro}Introduction}
The Tevatron has been a hotbed of \Bhad -physics activity since the
start of Run II in 2001. Although \ppb\ collisions at 1.96 TeV present
a much more challenging environment than that seen at the B-factories,
the fact that all types of \Bhad -hadrons are produced in CDF and
\Dzero\ makes 
the Tevatron \Bhad -physics program complementary to those of BaBar,
Belle, and CLEO. We will concentrate on two areas of this program:
measurements of \Bhad -lifetimes and the determination of the
oscillation frequencies between neutral \Bhad -mesons -- \Bhad
-mixing.
Beside their intrinsic interest, these two topics are representative
of the breadth of \Bhad -physics and highlight some of the key
experimental issues facing physicists attempting to study the
\qrk{b}-quark at hadron colliders.
Other topics in \Bhad -physics at the Tevatron are covered in
\cite{paulini,boudreau}.

The measurement of \Bhad -hadron lifetimes and of the neutral 
\Bhad -meson oscillation frequencies probe different aspects of the
Standard Model and its possible extensions. Lifetime measurements
provide input to our understanding of how to use the theory of
QCD. These measurements allow us to test extensions to the simple
spectator model of weakly decaying \Bhad -hadrons \cite{lenz}. In
particular, ratios of \Bhad -hadron lifetimes are sensitive to
different aspects of beyond-spectator-model effects and have now been
calculated to \Order{(\Lambda_{QCD}/m_b)^4)} \cite{lenz}.
\begin{eqnarray}
  \frac{\tau_1}{\tau_2} & = & 1 +
  \left( \frac{\Lambda_{QCD}}{m_b} \right)^2 \Gamma_2 \nonumber \\ 
  & & +
  \left( \frac{\Lambda_{QCD}}{m_b} \right)^3 
    \left[ \Gamma_3^{(0)} + \frac{\alpha_s}{4\pi} \Gamma_3^{(0)} +
    ... \right] \nonumber \\
  & & +
  \left( \frac{\Lambda_{QCD}}{m_b} \right)^4 
    \left[ \Gamma_4^{(0)} + ... \right]
\label{eqn:life}
\end{eqnarray}
In the above equation, the $\Gamma_2$ term is sensitive to
meson/baryon differences and
the $\Gamma_3$ terms reflect spectator-quark effects. Higher order
terms have been found to be negligible for ratios of \Bp , \Bd , and
\Bs\ mesons, but might be sizable for ratios involving \Bhad -baryons.
In the past, experimental results, particularly for
$\tau(\Lb)/\tau(\Bz)$,  have been in disagreement with
these expectations. As we will see, recent Tevatron results have gone
a long way towards clarifying the situation.

Lifetime results are also important inputs to the other topic covered
in these proceedings -- \Bhad -mixing measurements. Oscillations of neutral 
\Bhad -mesons (\Bd\ and \Bs ) are very sensitive to the mechanism of
electro-weak symmetry breaking because, within the Standard Model, the
CKM matrix, which describes quark mixing, is a consequence of the
Higgs mechanism. Since the Standard Model uses the simplest possible
method of electro-weak symmetry breaking -- a single Higgs doublet --
the CKM matrix is described, within this model, by only four
parameters -- three angles and a single \CP -violating phase. Other
models of particle physics are, in general, much less
constraining. Thus, measurements of the correlations between CKM
matrix elements provide powerful insight into physics beyond the
Standard Model.

Measurements of the frequency of \Bd\ and \Bs\ oscillations give us a
handle on the CKM matrix element, \Vckm{td}. The key to this
relationship lies in the different eigenstates of neutral \Bhad -meson
systems denoted {\it weak}, \CP , and {\it mass}. The time evolution
of the eigenstates of the weak interaction for \Bhad -mesons
containing $q=d,s$ quarks,
$| \Bzq{q}(t) \rangle$ and $| \Bzqbar{q}(t) \rangle$,
are governed by a Schr\"{o}dinger equation with off-diagonal elements:
\begin{equation}
\begin{array}{l}
  i \frac{\dderiv}{\dderiv t}
  \left( \begin{array}{c}
    | \Bzq{q}(t) \rangle \\
    | \Bzqbar{q}(t) \rangle
  \end{array} \right) \\
  =
  \left( \begin{array}{cc}
    M^{(q)} - i\Gamma^{(q)}/2 
    & M^{(q)}_{12} - i\Gamma^{(q)}_{12}/2 \\
    M^{(q)*}_{12} -  i\Gamma^{(q)*}_{12}/2 
    & M^{(q)} - i\Gamma^{(q)}/2
  \end{array} \right)
  \left( \begin{array}{c}
    | \Bzq{q}(t) \rangle \\
    | \Bzqbar{q}(t) \rangle
  \end{array} \right)
\end{array}
\end{equation}
Eigenstates of the system with definite mass,
$| B^H \rangle$ and $| B^L \rangle$,
are thus linear combinations of the weak eigenstates.
Oscillations between weak eigenstates then occur with a frequency
proportional to the mass difference:
\begin{equation}
  \Deltamq{q} = B^H_q - B^L_q \sim 2 | M_{12}^{(q)} |
\end{equation}
Further discussion of this, and of the \CP -eigenstates can be found
in \cite{boudreau}.

Although, in principle, only \Deltamq{d} is necessary to extract the
CKM matrix element, \Vckm{td}, in practice the theoretical uncertainty
on this extraction is reduced by more than a factor of three when the
ratio \dmd /\dms\ is used \cite{okamoto}:
\begin{equation}
  \frac{\dmd}{\dms} = 
  \left( \frac{M(\Bd )}{M(\Bs )} \right)
  \left( \frac{f^2_{B_d} B_{B_d}}{f^2_{B_s} B_{B_s}} \right)
  \left| \frac{\Vckm{td}}{\Vckm{ts}} \right| ^2
\label{eqn:vtdvts}
\end{equation}
Measurement of \dms\ has thus been a priority of the experimental high
energy physics community since the first measurements of \Bhad -mixing
\cite{bmix}.

\section{\label{sect:exp}Common Experimental Issues}
The busy environment surrounding a \Bhad -hadron produced in a \ppb\
collision at the Tevatron 
leads to  challenges for many aspects of the CDF and \Dzero\
detectors, particularly in the areas of triggering and tracking. These
challenges are apparent when considering the generic steps taken in a
lifetime or mixing analysis. First, candidate events must be recorded
from the 2.5 MHz beam collision rate. Then \Bhad -hadrons must be
reconstructed within the recorded events. Since both lifetime and
mixing analyses involve understanding the proper time evolution of 
\Bhad -decays, this quantity must be reconstructed by measuring 
the candidate
\Bhad -hadron's momentum and its decay length (the distance between the
production and decay points of the hadron). Background levels must
then be estimated. And finally, the relevant parameter ($\tau$ or
\Deltamq{}) must be extracted from a fit to the data of predictions
including all detector effects (efficiencies, resolutions, etc) and
background corrections.

Triggers are critical in the first step of this process. Both CDF and
\Dzero\ employ three-level trigger systems to reduce the intrinsic
interaction rate of $\sim$2.5 MHz (set by the bunch-crossing
frequency) to the 100-150 Hz of events that can be written to
permanent storage. The focus of the two experiments' \Bhad -physics
triggers is quite different though. CDF relies heavily on triggers
sensitive to tracks that are displaced from the primary vertex, while
\Dzero\ uses mainly single- and di-muon triggers to collect its 
\Bhad -physics sample. The difference in approach stems from the
different accept-rates allowed at the first level of triggering --
30 kHz for CDF compared to 2 kHz for \Dzero . Higher level-1 bandwidth
allows the CDF collaboration to collect a large sample of events
containing fully-hadronic \Bhad -hadron decays. \Dzero 's large
acceptance for muons, on the other hand, has allowed it to accumulate
a large sample of semi-muonic \Bhad -decays with little intrinsic
lifetime bias. As we will see, these types of triggers dictate the
type of analyses done by the two collaborations.

Once an event has been recorded, it is scrutinized for the presence of
\Bhad -hadrons. Because of its muon-based triggers, \Dzero\ tends to
start this process by identifying candidate semi-muonic 
\Bhad -decays while CDF uses a more inclusive approach. 
In the next steps,
tracking is used to identify charged particles potentially coming from
a \Bhad -hadron decay chain. Helpful in this process is the
reconstruction of intermediate states in the chain, such as
$D_s^- \ra \phi \pi^-$ and $\phi \ra K^+ K^-$. Unlike at the
B-factories, pion/kaon separation is not generally important here
(although CDF makes use of its d$E$/d$x$ and time-of-flight
capabilities in some analyses). Good invariant mass resolution is
critical, however, as reconstructed mass is generally used to identify
specific hadrons. CDF has the edge here (by nearly a factor of four)
because of their large-volume tracking system. Nevertheless, both
experiments are able to accumulate large samples of \Bhad -hadron
decays with high purities.

The final common feature in \Bhad -hadron reconstruction for lifetime
and mixing analyses is the 
estimation of the proper time of the \Bhad -hadron's decay. This
involves reconstructing the \Bhad 's production and decay points using
vertices found from combinations of charged tracks. Spatial resolution
of the tracking systems is thus a crucial element of lifetime and
mixing analyses. Both CDF and \Dzero\ have similar (and excellent)
capabilities here, with average uncertainties on proper time
reconstructed using only the charged particles from a \Bhad -decay of
around 50 $\mu$m for semi-leptonic \Bhad -decays and 25 $\mu$m for
fully hadronic decays. These resolutions are well below typical 
\Bhad -hadron lifetimes of $\sim$500 $\mu$m and are also smaller than
the \Bs\ oscillation period of $\sim$100 $\mu$m.

Reconstruction of the momenta of \Bhad -hadrons is, of course, also an
important element in estimating proper time. For the case of fully
reconstructed, hadronic \Bhad -decays the uncertainty that
this introduces in the estimate of proper time is
negligible compared to that associated with vertexing. For
semi-leptonic decays, however, the true \Bhad -hadron momentum cannot be
measured directly because of the presence of neutrinos in the decay. To
deal with this, correction factors, derived from simulation, are
applied to the reconstructed (from charged tracks) proper time of each
\Bhad -candidate based on its assumed flavor and mode.

\section{\label{sect:life}Lifetimes}

CDF and \Dzero\ play an important role in our understanding of the
lifetimes of weakly decaying \Bhad -hadrons. Not only are these
experiments the only place where higher mass \Bhad -mesons and all
\Bhad -baryons can be studied, they also provide competitive
lifetime results for \Bz\ and \Bp\ mesons. 
In that area, CDF has produced recent preliminary
measurements using 
fully reconstructed \Bz\ and \Bp\ decays to
$D^{0/+}$ mesons and charged pions;
semi-leptonic decays involving $D$ and $D^*$ mesons;
and decays to \Jpsi $K^{(*)}$.
\Dzero\ has a published result on the \Bp /\Bz\ lifetime ratio using
$\mu D^{(*)} X$ final states.
Taken together, these Tevatron results have a weight of $\sim$38\% in
the $\tau(\Bp )/\tau(\Bz )$ world average \cite{hfag07}.

The Tevatron has also been active in studies of the \Bs -meson, as we
have seen in several contributions to this conference
\cite{boudreau}. The \Bs\ lifetime plays a
foundational role in these studies and both CDF and \Dzero\ have
measured this quantity in a variety of ways. Care must be taken when
interpreting \Bs\ lifetime results as these mesons have a
non-negligible width difference between their mass
eigenstates \cite{boudreau}, which, 
for these purposes, are approximately equivalent to the \CP\
eigenstates. \Bs\ lifetime results are thus only given for those
decays in which the \CP\ content is well-known and are quoted as the
average lifetime of the heavy and light mass eigenstates.
CDF and \Dzero\ have several new results in this area, summarized in
Table~\ref{table:life}.

\begin{figure}[h]
\centering
\TwoColFig{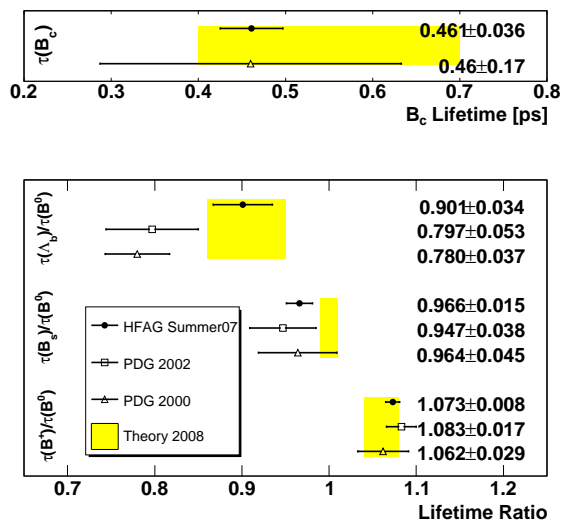}
\caption{A comparison of world average lifetime ratios, and \Bc\
  lifetimes in 2000, 2002, and 2008 with current theoretical predictions.}
\label{fig:life}
\end{figure}

Measurements of \Bhad -baryon properties are also an important
component of the Tevatron \Bhad -physics program
\cite{paulini}. Although several new baryon states have recently been
observed by CDF and \Dzero\ \cite{paulini}, sufficient statistics to
make a lifetime determination have only been accumulated for the 
\Lb -baryon. Both \Jpsi $\Lambda$ and $\mu \Lambda_c$ final states
have been studied. Results are summarized in
Table~\ref{table:life}. They indicate generally good agreement with
theoretical expectations (see Fig.~\ref{fig:life}). However, there is
some discrepancy between the CDF and \Dzero\ results that will need to
be understood with more data.

The last lifetime measurement we will mention, that of the \Bc -meson,
probes different theoretical issues than those states discussed
previously. Because the \Bc\ is composed of two heavy quarks, it has
more spectator-level decay possibilities than other weakly decaying
\Bhad -hadrons. In fact, theory predicts that the \Bc\ lifetime should
be approximately one third of that of the other \Bhad -mesons.

On the experimental side, both CDF and \Dzero\ have now collected
large samples of \Bc\ candidates in semi-leptonic decay modes, as well
as smaller sets of fully hadronic decays. 
Only semi-leptonic decays are currently used
for the determination of the \Bc\ lifetime, results of which are
summarized in Table~\ref{table:life}.

\begin{table}[h]
\begin{center}
\caption{Recent \Bhad -lifetime measurements from CDF and \Dzero
  . Integrated luminosity is given in \unitexp{fb}{-1}.}
\begin{tabular}{|ll|c|c|c|c|}
\hline
  \textbf{Mode} & \textbf{Exp} & \textbf{Lumi}
  & \textbf{Signal} & \textbf{$\tau$ (ps)} \\
\hline
  \Bs \ra \Jpsi $\phi$ & CDF & 1.7 & 2500
    & 1.52$\pm$0.04$\pm$0.02 \cite{cdfjpsiphi} \\
  & \Dzero & 2.8 & 1976
    & 1.52$\pm$0.05$\pm$0.01 \cite{d0jpsiphi} \\
  \Bs \ra $\pi$\Ds & CDF & 1.3 & 3340
    & 1.517$\pm$0.041$\pm$0.025 (prel) \\
  \Bs \ra $\ell$\Ds $X$ & CDF & 0.4 & 1155
    & 1.381$\pm$0.055$^{+0.052}_{-0.048}$ (prel) \\
  & \Dzero & 0.4 & 5176
    & 1.398$\pm$0.044$^{+0.028}_{-0.025}$ \cite{d0bslx} \\
\hline
  \Lb \ra \Jpsi $\Lambda$ & CDF & 1.0 & 557
    & 1.58$\pm$0.08$\pm$0.01 (prel) \\
  & \Dzero & 1.2 & 171
    & 1.22$^{+0.13}_{-0.11}\pm$0.04 \cite{d0lbjpsil} \\
  \Lb \ra $\mu \Lambda_c X$ & \Dzero & 1.3 & 3727
    & 1.29$^{+0.12}_{-0.11}\pm$0.09 \cite{d0lbmulcx} \\
\hline
  \Bc \ra \Jpsi $\ell$ & CDF & 1.0 & 916
    & 0.475$^{+0.053}_{-0.049}\pm$0.018 (prel)\\
  \Bc \ra \Jpsi $\mu$ & \Dzero & 1.3 & 881
    & 0.448$^{+0.038}_{-0.036}\pm$0.032 \cite{d0bcjpsimu}\\
\hline
\end{tabular}
\label{table:life}
\end{center}
\end{table}

As can be seen in Fig. \ref{fig:life}, Run II Tevatron measurements of
\Bhad -hadron lifetimes have dramatically increased the precision with
which we can probe QCD. In this figure we see a comparison of world
average results from 2000 \cite{pdg2000} and 2002 \cite{pdg2002}
(before Run II results) and the current world averages from HFAG
\cite{hfag07}.
Also included are theoretical predictions of 
lifetime ratios \cite{thliferat} and \Bc\ lifetimes \cite{thbclife}.
The latest experimental measurements represent improvements of factors of
three over pre-Run II results. They tend to be in good agreement with
theoretical expectations. In some cases, experimental precision is
smaller than uncertainties in the calculations allowing constraints to
be put on models of \Bhad -decays.

\section{\label{sect:mix}Mixing}

Measurement of the oscillation frequency between \Bzq{s} and
\Bzqbar{s} mesons (or equivalently, the mass difference, \dms ,
between the heavy and light mass eigenstates) was one of the goals of
Run II at the Tevatron. This goal was achieved in the spring of 2006
when \Dzero\ saw first hints of a mixing signal \cite{d0dms}, followed
quickly by a $>$3$\sigma$ significance measurement by CDF
\cite{cdfdmsev}. These early results have now been updated with a new
published measurement from CDF \cite{cdfdmsmeas} and a set of preliminary
results from \Dzero . Basic features of the analyses are given in
Table~\ref{table:dms}, and are described in more detail in the following.

\begin{table}[h]
\begin{center}
\caption{A comparison of \Bs -candidate sample sizes, tagging power
  (\eDsq , for opposite and same side tagging), and sensitivity
  between various decay modes used in the CDF \cite{cdfdmsmeas} and
  \Dzero\ preliminary analyses 
  and in the previously most sensitive result from
  ALEPH \cite{alephdms}.}
\begin{tabular}{|ll|r|cc|r|}
\hline
  & & & \multicolumn{2}{|c|}{$\mathbf{\varepsilon D^2}$} 
    & \textbf{Sensitivity}\\
  \textbf{Exp} & \textbf{Mode} & \textbf{Sample}
    & \textbf{OST} & \textbf{SST} & \textbf{(\unitexp{ps}{-1})}\\
\hline
  ALEPH & Hadronic & 28.5 & \multicolumn{2}{|c|}{27\%} & 13.6\\
\hline
  \Dzero & $\ell$\Ds & 64,500 & \multicolumn{2}{|c|}{4.5\%} & 25.4\\
  & $\pi$\Ds & 249 & 2.5\% & & 14.0\\
\hline
  CDF & $\ell$\Ds & 61,500 & 1.8\% & 4.8\% & 19.3\\
  & (3)$\pi$\Ds & 8,700 & 1.8\% & 3.7\% & 30.7\\
\hline
\end{tabular}
\label{table:dms}
\end{center}
\end{table}

The analyses producing the results mentioned above are similar to
lifetime analyses in that they examine the proper time evolution of
candidate \Bs\ decays. An added feature is the use of {\it tagging} to
determine the flavor (\Bzq{s} or \Bzqbar{s}) of the meson at
production and decay. Tagging the decay flavor is straightforward
using charges of the decay products. Production flavor tagging,
however, is harder. Two classes of techniques are used here: opposite
side tagging (OST) and same side tagging (SST).

Because \qrk{b}-quarks are produced in quark-antiquark pairs at the
Tevatron, the OST technique uses information about the ``other''
(non-\Bs ) \Bhad -hadron in the event to determine its production
flavor. The \Bs\ candidate is then assumed to have been produced with
the opposite flavor. SST, on the other hand, uses information gleaned
from fragmentation and other particles associated with the \Bs -meson
itself to determine its flavor at production. The figure of merit
associated with these techniques is called the {\it tagging power},
\eDsq . It is composed of the efficiency, $\varepsilon$, for an event to
be tagged as either oscillated or non-oscillated and the dilution,
$D$, a quantity related to the purity of the tagging method:
$D = 2\eta - 1$, where $\eta$ is the fraction of tags where the
oscillated or non-oscillated state is correctly identified. Tagging
powers for the various modes used in CDF and \Dzero\ \Bs\ mixing
analyses are give in Table~\ref{table:dms}.

Understanding tagging is obviously a critical component of mixing
analyses into which the experiments have put a lot of effort. Briefly
speaking, OST is calibrated by measuring the well-known \Bd\
oscillation frequency. The \Bd\ mass difference, \dmd , has been
measured extremely accurately by BaBar and Belle (see \cite{hfag07}
for a compilation of these results). CDF and \Dzero\ \cite{d0dmd}
cannot compete with these measurements, but do use \dmd\ analyses to
simultaneously determine OST calibration parameters. The results
obtained for \dmd\ are fully consistent with the world average, giving
us confidence in the OST technique.

Same side tagging performance, unlike that of the OST, is dependent
upon the particular \Bhad -meson flavor being considered. The method
to verify its calibration therefore consists of ensuring that the SST
gives similar results in data and MC control samples, such as
\Bp \ra \Jpsi $K^+$. The MC is then assumed to give a correct
description of the SST in \Bs -events.

\begin{figure*}[t]
\centering
\TwoColFig{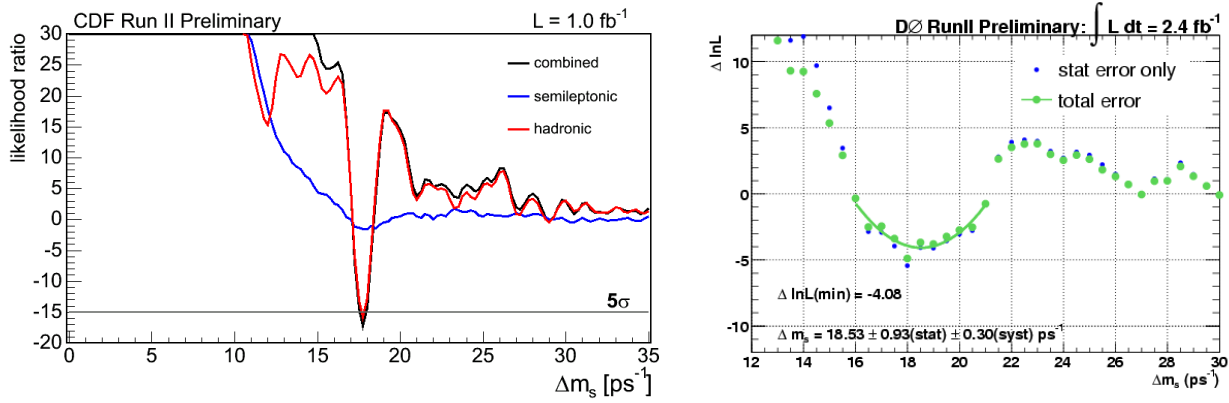}
\caption{Scans of $-\ln \mathcal{L}$ vs \dms\ for the CDF (left plot) and
  \Dzero\ (right plot) \Bs\ mixing analyses.}
\label{fig:dms}
\end{figure*}

With tagging well in hand, fits are performed to determine the value
of \dms . Scans of $\ln \mathcal{L}$ vs. assumed \dms\ are shown in
Fig.~\ref{fig:dms} for all analysis modes in CDF
\cite{cdfdmsmeas} and for the preliminary, combined  \Dzero\ result. 
Both
experiments see minima in $-\ln \mathcal{L}$ around 18
\unitexp{ps}{-1}, measuring the following values for \dms :
\begin{eqnarray}
  17.77 \; \pm \; 0.10 \; \pm \; 0.07 \unitexp{ps}{-1}
    & \mathrm{CDF} & \mbox{\cite{cdfdmsmeas}} \nonumber \\
  18.53 \; \pm \; 0.93 \; \pm \; 0.30 \unitexp{ps}{-1}
    & \Dzero & \mathrm{(prelim)}
\end{eqnarray}

Scans of $-\ln \mathcal{L}$ vs \dms\ for the two experiments are shown
in Fig.~\ref{fig:dms}.
The significance of the CDF result is 5.4$\sigma$ (background
fluctuation probability of \scinot{8}{-8}), while that of the \Dzero\
measurement is 2.9$\sigma$ (with systematic effects included).

Using the two measurements above and Eq.~\ref{eqn:vtdvts}, we find an
average value of $|\Vckm{td} / \Vckm{ts}|$ of:
\begin{equation}
  \left| \frac{\Vckm{td}}{\Vckm{ts}} \right| =
  0.2060 \; \pm 0.0012 \mathrm{(exp)} 
  \; ^{+0.0081}_{-0.0060}\mathrm{(theor)}
\end{equation}
where the ``exp'' error includes all statistical and systematic errors
on the measurements of \dms , while the ``theor'' error comes from the
uncertainty on the ratio of decay constants and bag parameters from
lattice calculations \cite{okamoto}:
\begin{equation}
  \xi \equiv
  \frac{f_{Bs} \sqrt{B_{Bs}}}{f_{Bd} \sqrt{B_{Bd}}} =
  1.210^{+0.047}_{-0.035}
\end{equation}

It is interesting to note that the theoretical error on $\xi$
completely dominates the uncertainty of the CKM element ratio (by a
factor of nearly 7) and also that the relative error on the world
average value of \dms\ (0.3\%) is now smaller than that on \dmd\
(0.5\%) \cite{hfag07}.

\begin{figure*}[t]
\centering
\TwoColFig{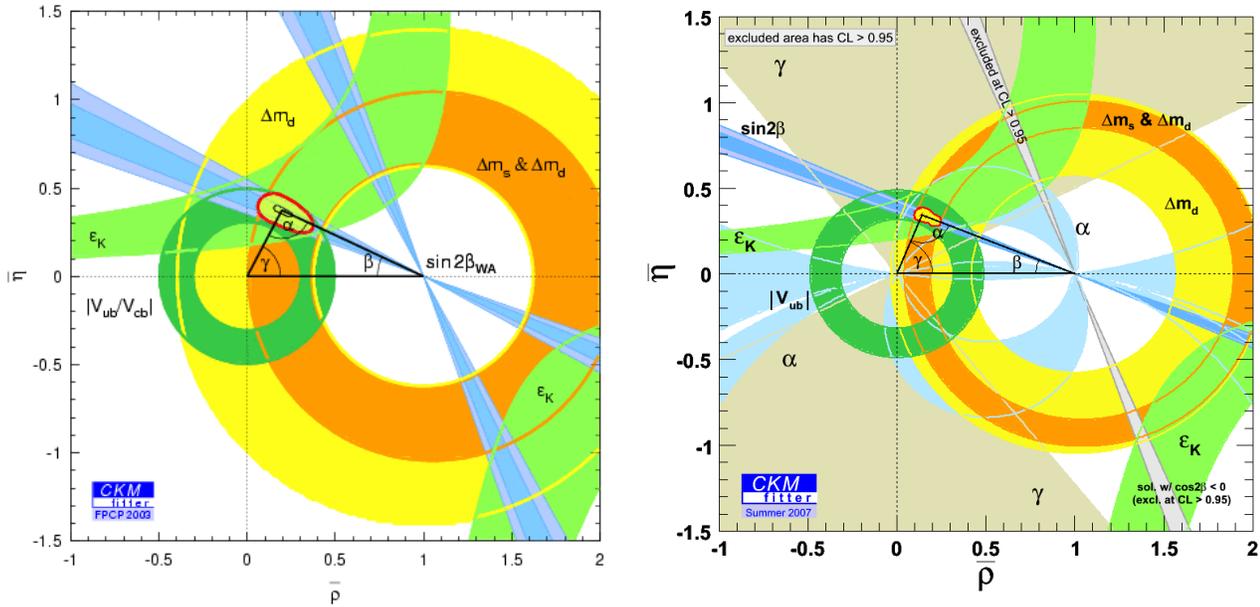}
\caption{Experimental constraints on the unitarity triangle in 2003
  (left plot) and 2007 (right plot) from the CKMfitter group.}
\label{fig:ut}
\end{figure*}

These new measurements of \dms\ have a large impact on tests of the
consistency of the CKM picture of quark mixing. This is particularly
evident when comparing experimental reconstruction of the ``unitarity
triangle'' (formed from one element of the CKM unitarity condition:
\Vckm{ud}\Vckmst{ub}+\Vckm{cd}\Vckmst{cb}+\Vckm{td}\Vckmst{tb}=0).
Figure~\ref{fig:ut} shows a comparison of the state of our knowledge
of this triangle in 2003 and 2007 \cite{ckmfitter}
(see also \cite{utfit}).
The new measurement of \dms\ significantly increases the accuracy with
which we know the apex of the unitarity triangle and favors those
models of new physics that have a Standard Model, CKM-like flavor
structure. 

\section{\label{sect:concl}Conclusions and Future Prospects}
We've made remarkable progress in the field of \Bhad -physics since
the start of Run II at the Tevatron. 
To cite just a few examples discussed here:
our knowledge of \Bhad -hadron lifetimes has improved by a factor of
two or more depending on the hadron; 
the accuracy to which we measure
the \Bd\ mixing frequency has increased by a factor of more than
three, thanks to hard work at the B-factories;
and we have finally observed \Bs\ oscillations, which are now measured
with an uncertainty of only 0.3\%.
All these measurements, and many others, point to a picture of flavor
that is consistent with the Standard Model CKM description. However,
some cracks may be appearing in the mirror due to recent measurements
of \CP\ violation in \Bs -mesons \cite{boudreau}.

Certainly then, the \Bs\ is a system to watch --
particularly in the \CP -sector.
Advances on the oscillation frequency side will need to come from
improved calculations though, since that is where the main source of
uncertainty now lies.
For lifetime measurements, the increasingly large data sets being
collected at the Tevatron (more than 4 \unitexp{fb}{-1} have now been
delivered to each experiment) hold out the prospect of comprehensive tests
of QCD models using a wide range of \Bhad -hadrons. The experimental
focus here will shift to baryons (and the \Bc ), 
including the newly observed $\Xi_b$
and $\Sigma_b$ states.

With many past successes and a bright future ahead,
\Bhad -physics will remain a vital part of the Tevatron program while
we await the next big step -- LHCb.

\begin{acknowledgments}
I am pleased to acknowledge the assistance of my collaborators on
\Dzero\ and CDF in the preparation of this presentation. I would also
like to thank the HQL08 organizers for a very interesting and
enjoyable conference!
\end{acknowledgments}

\bigskip 

\end{document}